\documentclass[fleqn,usenatbib]{mnras}

\usepackage{newtxtext,newtxmath}

\usepackage[T1]{fontenc}

\DeclareRobustCommand{\VAN}[3]{#2}
\let\VANthebibliography\thebibliography
\def\thebibliography{\DeclareRobustCommand{\VAN}[3]{##3}\VANthebibliography}

\usepackage{graphicx}	
\usepackage{amsmath}	
\usepackage{booktabs}
\usepackage{siunitx}
\usepackage{gensymb}
\usepackage{xcolor}

\title[Updated densities of the young V1298 Tau system]{CHEOPS observations of V1298 Tau: updated planetary densities and implications on the early evolution of the young system}

\author[H. Shivkumar et al.]{
Hinna Shivkumar,$^{1}$\thanks{E-mail: h.shivkumar@uva.nl}
Sérgio Gomes,$^{1}$
Jean-Michel Désert,$^{2,3,1}$
John Livingston,$^{4,5,6}$
John Lopez,$^{1}$
\newauthor
Vatsal Panwar,$^{7}$
James Sikora,$^{8}$
Saugata Barat,$^{9,10}$
Pierre F. L. Maxted,$^{11}$
Thomas G. Wilson,$^{12}$
\newauthor
Antonija Oklopčić,$^{1}$
Silvia Toonen$^{1}$
\\
$^{1}$Anton Pannekoek Institute for Astronomy, University of Amsterdam, Science Park 904, 1098 XH, Amsterdam, The Netherlands\\
$^{2}$Leibniz-Institut für Astrophysik Potsdam (AIP), An der Sternwarte 16, 14482 Potsdam, Germany\\
$^{3}$DESY, Platanenallee 6, D-15738 Zeuthen, Germany\\
$^{4}$Astrobiology Center, Mitaka, Japan.\\
$^{5}$National Astronomical Observatory of Japan, Mitaka, Japan\\ $^{6}$Department of Astronomical Science, The Graduate University for Advanced Studies, Hayama, Japan\\
$^{7}$School of Physics \& Astronomy, University of Birmingham, Edgbaston, Birmingham B15 2TT, UK\\
$^{8}$Lowell Observatory, 1400 W Mars Hill Road, Flagstaff, AZ, 86001, USA\\
$^{9}$Kavli Institute for Astrophysics and Space Research, Massachusetts Institute of Technology, Cambridge, MA 02139, USA\\
$^{10}$University of Southern Queensland, West St, Darling Heights, Toowoomba, Queensland, 4350, Australia\\
$^{11}$Astrophysics Group, Keele University, Staffordshire ST5 5BG\\
$^{12}$Department of Physics, University of Warwick, Coventry CV4 7AL, UK
}

\date{Accepted XXX. Received YYY; in original form ZZZ}

\pubyear{\the\year{}}

\begin{document}
\label{firstpage}
\pagerange{\pageref{firstpage}--\pageref{lastpage}}
\maketitle

\begin{abstract}
The young (10-30 Myr) multi-planet system V1298 Tau presents a unique opportunity to probe the early formation and evolution of young systems. We present new CHEOPS observations of the three innermost planets, yielding high-precision planetary radii ($\sim$5-11 R$_\oplus$) and improving the radius ratios ($\rm R_{\mathrm{p}}/R_{\mathrm{s}}$) by 30-71\% compared to previous multiple TESS observations. Combined with refined period and mass determinations from transit-timing variation (TTV) measurements, we derive revised bulk densities (0.06-0.23 $\rm g/cm^3$) for these planets. We find that the innermost planet c is denser compared to the outermost planet at the 3.4-$\sigma$ level, while the bulk densities of the three outermost planets are consistent within the reported uncertainties. These bulk densities suggest differing envelope mass fraction across the system, indicating differential atmospheric evolution in the young system. We further assess the early dynamical state of the V1298 Tau system and find that within the range of simulations performed we find no evidence for present-day mean-motion resonance trapping. As an independent diagnostic, we compute the forced eccentricities and low Normalized Angular Momentum Deficit (NAMD) exhibited by the system. Our simulations suggest that no past dynamical excitation is required to explain the present orbital architecture.
\end{abstract}




\section{Introduction}


The evolution of the orbital architectures of planetary systems during their early stages remains poorly understood. This is largely because the physical parameters that govern orbital migration and tidal dissipation are weakly constrained for young planets. Population studies reveal that resonant chains are more prevalent in young systems but become increasingly rare with age \citep{Dai_etal_2024}. This trend suggests that resonance capture is a natural outcome of planet formation and disk-driven migration, yet the mechanisms that maintain and subsequently disrupt these chains, from divergent migration to orbital instabilities, remain uncertain \citep[e.g.][]{Liu_etal_2022, Hansen_etal_2024, Pu_Wu_2015}. 

A central challenge lies in the limited precision of planetary radii and masses of young planets. These parameters strongly influence the strength of tidal interactions. Tidal dissipation is particularly sensitive to planetary size, with theoretical models indicating a steep dependence, scaling as $\dot{a}_i \propto R_i^5$ \citep[e.g. 
Eq. 20,][]{Correia_etal_2011}. As a result, fractional uncertainties in radius translate to large uncertainties in the migration rate and, consequently, the inferred tidal evolution. Similarly, precise planetary masses are essential for determining the tidal response since the resulting tidal migration rates also depend sensitively on the mass, scaling as $\dot{a}_i \propto m_i^2$. Without high-precision measurements of both radius and mass of young planets, constraints and predictions of early dynamical evolution remain fundamentally limited.

Young planetary systems ($<$100  Myr) provide a distinctive dynamical environment for studying their early dynamical states. Yet, their characterization is often hindered by high stellar activity and limited precision in photometry. The CHaracterising ExOPlanet Satellite \citep[CHEOPS;][]{Benz_2021} is uniquely suited to overcome these limitations, using its ability to obtain high-precision targeted observations of known transiting planets. This enables CHEOPS to provide substantial improvements in radius precision, even around photometrically variable young stars. When combined with complementary mass refinement efforts, such as transit-timing variations (TTVs) and atmospheric observations, CHEOPS radii can place young planets in a powerful position to constrain early dynamical evolution. 

\begin{table*}
\centering
\caption{Log of CHEOPS observations of V1298 Tau with start and end times, duration of observations, exposure time, number of data points, observing efficiency, and planet(s) observed during each visit.}
\label{tab:obslog}
\begin{tabular}{lccccccc}
\toprule
Visit &
Start Time &
End Time &
Duration &
Exposure Time &
Data Points &
Efficiency &
Transit(s) Observed \\
 & (UTC) & (UTC) & (h) & (s) & (Non-flagged) & (\%) &  \\
\midrule
1 & 2021-12-03T19:49:03 & 2021-12-04T20:25:20 & 24.60 & 30.0 & 1733 & 58.7 & V1298 Tau d \\
2 & 2021-12-14T19:22:23 & 2021-12-15T16:26:29 & 21.07 & 30.0 & 1487 & 58.8 & V1298 Tau c \\
3 & 2021-12-28T13:29:43 & 2021-12-29T14:06:00 & 24.60 & 30.0 & 1632 & 55.3 & V1298 Tau b, d \\
4 & 2022-01-21T17:06:42 & 2022-01-22T17:02:27 & 23.93 & 30.0 & 1542 & 53.7 & V1298 Tau b \\
\bottomrule
\end{tabular}
\end{table*}

V1298 Tau \citep[9-30 Myr;][]{david_2019b, finociety_2023} has emerged as a benchmark among the growing library of such young systems discovered to date \citep[e.g.,][]{david_2019b, david_2019a, benatti_2019, plavchan_2020, newton_2021}. Discovered during the \textit{K2} Campaign 4 \citep{howell_2014}, V1298 Tau is a bright (V$\approx$10), pre-main sequence weak-lined T Tauri star hosting four confirmed transiting planets \citep{david_2019b, david_2019a, Feinstein_etal_2022}, spanning between 5-10 $R_\oplus$. \citet{livingston_2026} refined the masses and orbital periods (8-48 days) of the V1298 Tau planets using TTV analysis. This places the masses of the planets between 5-15 $M_\oplus$, with the mass of V1298 Tau b in agreement with atmospheric mass determination from \citep{barat_2025}. The TTV analysis also finds low eccentricities for the planets in the system (e$\sim$0.01) as opposed to larger eccentricities (e=0.1) reported previously \citep[e.g.,][]{Suarez_etal_2022, sikora_2023}.

\cite{Turrini_etal_2023} found strong evidence of past dynamical excitation in the V1298 Tau system, provided by the normalized angular momentum deficit (NAMD) calculations. \citep{Tejada_etal_2022} demonstrated the planets' proximity to the resonance chains 3:2, 2:1, 3:2 and 3:2, 2:1, 2:1, further supporting the hypothesis of an earlier resonant chain configuration. However, these studies used planetary masses retrieved from radial velocity measurements from \citep{Suarez_etal_2022}, which were reported by \citet{blunt_2023} to have over-predicted planet masses due to overfitting. The large uncertainties in planetary radii and  masses combined with unconstrained period of the outermost planet V1298 Tau e have previously limited efforts to robustly predict the system's early tidal evolution and dynamical history.

In this paper, we present high-precision CHEOPS transit observations of the three innermost planets (c, d and b) of the young V1298 Tau system and significantly improve the precision on their radius measurements and mid-transit times. These refined radii, when combined with updated masses (5-15 M$_\oplus$) and periods from TTV analysis \citep{livingston_2026} allow us to better evaluate the dynamical evolution of the V1298 Tau system and assess if the system is currently trapped in a mean-motion resonance (MMR).

The paper is organized as follows. In Section \ref{sec:obs}, we describe the CHEOPS observations. In Section \ref{sec:DA}, we describe the data analysis techniques and transit fits for each CHEOPS observation. We present the results in Section \ref{sec:Res}. In Section \ref{sec:disc}, we assess if the V1298 Tau planets are or have been in a mean-motion resonance.

\begin{figure*}
    \centering
    \includegraphics[width=1.0\textwidth]{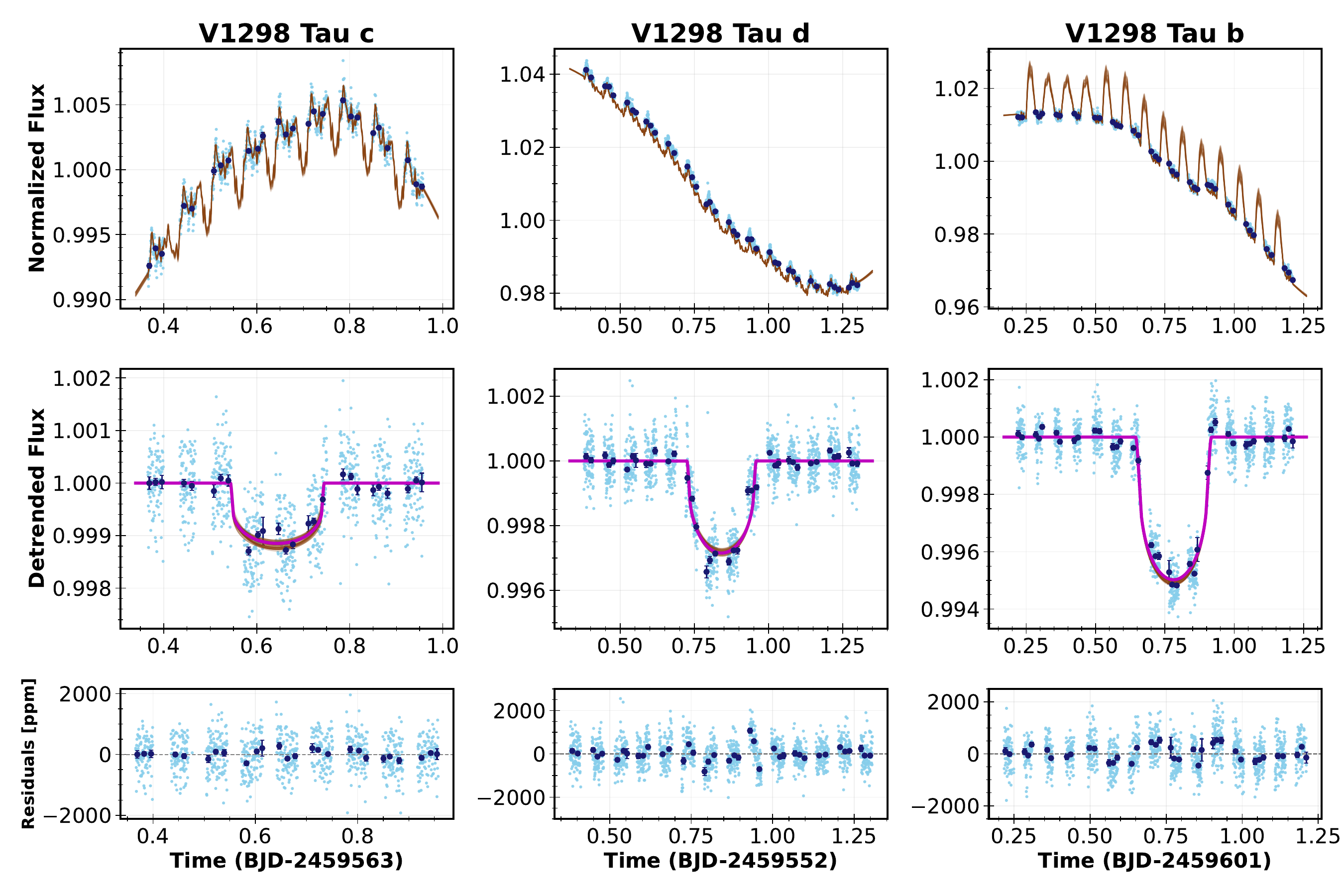}\hfill
    \caption{Individual CHEOPS transit observations of V1298 Tau c, d, and b displayed in the left, middle, and right panels, respectively. For each planet, the top row presents the raw (pre-detrended) light curves, the middle row shows the corresponding detrended light curves, and the bottom row shows the residuals from the best-fit model. Unbinned data are shown in sky blue, binned data in dark blue, and the best-fit transit model in magenta. The solid brown lines indicate 50 random samples from the MCMC posterior distribution of the combined transit and systematics model.}
    \label{fig:cheops_lcs}
\end{figure*}

\begin{table*}
\centering
\caption{Overview of the best-fit transit parameters for V1298 Tau c, d, and b using \textsc{pycheops}.}
\label{tab:fit_parms}
\begin{tabular}{lccc}
\toprule
Parameter [units] & Symbol & Prior & Best-fit value \\
\midrule
\multicolumn{4}{c}{V1298 Tau c} \\
\midrule
Mid-transit time [BJD$_{\mathrm{TDB}}$ $-$ 2459563] & T$_{0,c}$ & $\mathcal{N}$(0.644, 0.05) & $0.64399 \pm 0.00088$ \\
Orbital period [days] & P$_c$ & fixed & 8.24872 \\
Transit depth [ppm] & D$_c$ & $\mathcal{U}$(313, 1253, 5012) & $965 \pm 53$ \\
Transit duration [days/orbital period] & W$_c$ & fixed & 0.0244 \\
Impact parameter & b$_c$ & fixed & 0.14 \\
Limb-darkening coefficient & h$_1$ & $\mathcal{N}$(0.675, 0.10) & $0.736 \pm 0.053$ \\
Limb-darkening coefficient & h$_2$ & fixed & 0.434 \\
Radius ratio [stellar radii] & R$_p$/R$_s$ & -- & $0.03106 \pm 0.00090$ \\
Scaled semi-major axis & $a/R_s$ & -- & $13.34 \pm 0.012$ \\
\midrule
\multicolumn{4}{c}{V1298 Tau d} \\
\midrule
Mid-transit time [BJD$_{\mathrm{TDB}}$ $-$ 2459552] & T$_{0,d}$ & $\mathcal{N}$(0.822, 0.05) & $0.84233 \pm 0.00086$ \\
Orbital period [days] & P$_d$ & fixed & 12.40214 \\
Transit depth [ppm] & D$_d$ & $\mathcal{U}$(460, 1840, 7362) & $2160 \pm 55$ \\
Transit duration [days/orbital period] & W$_d$ & fixed & 0.0188 \\
Impact parameter & b$_d$ & fixed & 0.19 \\
Limb-darkening coefficient & h$_1$ & $\mathcal{N}$(0.675, 0.01) & $0.6301 \pm 0.0088$ \\
Limb-darkening coefficient & h$_2$ & $\mathcal{N}$(0.434, 0.01) & $0.463 \pm 0.050$ \\
Radius ratio [stellar radii] & R$_p$/R$_s$ & -- & $0.04647 \pm 0.00062$ \\
Scaled semi-major axis & $a/R_s$ & -- & $17.42 \pm 0.011$ \\
\midrule
\multicolumn{4}{c}{V1298 Tau b} \\
\midrule
Mid-transit time [BJD$_{\mathrm{TDB}}$ $-$ 2459601] & T$_{0,b}$ & $\mathcal{N}$(0.791, 0.05) & $0.78114 \pm 0.00049$ \\
Orbital period [days] & P$_b$ & fixed & 24.14041 \\
Transit depth [ppm] & D$_b$ & $\mathcal{U}$(1132, 4529, 18117) & $4103 \pm 63$ \\
Transit duration [days/orbital period] & W$_b$ & fixed & 0.0113 \\
Impact parameter & b$_b$ & fixed & 0.45 \\
Limb-darkening coefficient & h$_1$ & $\mathcal{N}$(0.675, 0.01) & $0.640 \pm 0.011$ \\
Limb-darkening coefficient & h$_2$ & $\mathcal{N}$(0.424, 0.05) & $0.553 \pm 0.060$ \\
Radius ratio [stellar radii] & R$_p$/R$_s$ & -- & $0.06386 \pm 0.00049$ \\
Scaled semi-major axis & $a/R_s$ & -- & $27.16 \pm 0.02$ \\
\bottomrule
\end{tabular}
\end{table*}

\section{Observations} \label{sec:obs}
In this paper, we used time-series observations from CHEOPS photometry (Section \ref{sec:obs_cheops}).

\subsection{CHEOPS Photometry} \label{sec:obs_cheops}
We observed transits of V1298 Tau c, d, and b with CHEOPS as part of the GO program ID 0010 (PI: Désert). The program consisted of four visits of V1298 Tau between 03 December 2021 and 21 January 2022. Individual transits of V1298 Tau d, c, and b were observed during the first, second, and fourth visits, respectively, while the third visit captured a joint transit of V1298 Tau d and b. Each CHEOPS visit lasted between 21.1 and 24.6 hours to encompass the full transit and to ensure sufficient out-of-transit baseline while also accounting for the large TTVs exhibited by the system. Gaps within the data due to Earth occultation and passage over the South Atlantic Anomaly (SAA) limited the observing efficiency of the observations to 53-59\%. The exposure time for all observations was set to 30.0\,s. A summary of the observation log is shown in Table \ref{tab:obslog}.

By default, the raw CHEOPS images were processed with the CHEOPS Data Reduction Pipeline (DRP; version 13.1.0; \cite{Hoyer2020}), which performs image calibration, systematic corrections, and aperture photometry to extract the light curves. The PSF of V1298 Tau is primarily contaminated by the extended PSF of a fainter star (\textit{Gaia} DR3 51886331671984640; m$_{Gaia} \simeq$ 16.8824). Standard aperture photometry can partially clip the outer PSF of V1298 Tau to mitigate contamination from nearby stars, partially removing target flux as well. To minimize these losses, we applied PSF photometry using the PSF Imagette Photometric Extraction (PIPE) tool \citep{Brandeker_2024}. PIPE employs principal component analysis (PCA) to model the target and background PSFs by searching a library of reference PSFs for the best-fit match using key parameters such as target coordinates, stellar effective temperature, and telescope temperature. We provide details of the PIPE reduction for our CHEOPS observations in Appendix A.1.

\section{Data Analysis}\label{sec:DA}
\subsection{CHEOPS light curve fitting}
The CHEOPS light curves of V1298 Tau extracted with PIPE were analyzed using \verb|pycheops| \citep{maxted2021}. Since PSF photometry removes flux modulation from nearby stars, we skip the decontamination method in \verb|pycheops| which is generally used for light curves extracted using aperture photometry with DRP. For the light curve fitting, we simultaneously fit for the transit model, telescope systematics, and host-star variability. 

For the analysis, the light curves are normalized to their median. We applied consistent fitting routines to each individual CHEOPS transit of  V1298 Tau c, d, and b, unless otherwise noted. CHEOPS light curves are affected by instrumental systematics. For instance, the flux measurements are known to be correlated with the roll angle of the telescope, primarily linked to the orbital orientation of CHEOPS. To model and correct these effects, we used \verb|pycheops| for model selection and systematics detrending, following the procedure described by \citet{maxted2021}. 

For model selection, we use the Bayes Factor to determine the optimal set of decorrelation parameters for instrumental systematics and stellar baseline correction. We implement this by adding decorrelation parameters one by one and selecting the parameters with the lowest Bayes Factor and stopping when B$_p>1$ for all remaining parameters. The model selection process is described in detail in \citet{maxted2021}. 

The detrending model that we consider includes decorrelation parameters defining the instrumental systematics such as the first- and second-order derivatives of the centroid offset in x and y coordinates, the first and second harmonics of the spacecraft roll angle, and background flux variation. We do not include contamination and smearing terms since these have been corrected for in the PIPE reduction. 

The detrending model also includes decorrelation parameters to account for the long timescale flux variation of V1298 Tau (e.g. the CHEOPS visit of V1298 Tau d shows an overall change in flux of 7\%). We include higher order baseline models (up to 5$^{\rm th}$-order polynomial) in \verb|pycheops| to model the stellar variability. The polynomial order for the baseline is selected during the model selection process described above. The final set of decorrelation parameters used to fit each observation are listed in Table \ref{tab:decorrparms}. The fit to the trends in roll angle for each observation are shown in Figure \ref{fig:rollangle}.

Following \citet{maxted2021}, we first performed an initial fit using the Levenburg-Marquardt algorithm for non-linear least-squares minimization, implemented using the \verb|LMFIT| package \citep{newville_2014}. For these preliminary fits, we adopted orbital and planetary parameters from \citep{sikora_2023} as priors to determine the mid-transit time. Then, the residual RMS from this fit was used to perform model selection and set the widths for the normal distribution of the priors on the selected decorrelation parameters. Then, for each visit, we fit the transit model simultaneously using the decorrelation priors and baseline parameters with \verb|LMFIT|. For the transit model, we fit the transit depth, mid-transit time, and limb-darkening coefficients. The best-fit values from the \verb|LMFIT| are used as priors in a Markov Chain Monte Carlo (MCMC) \citep{foreman_mackey_2013} to derive posterior distributions for the fitted parameters. The orbital period, transit duration, impact parameter were fixed to literature values of \citet{sikora_2023}. Assuming a circular orbit, the eccentricity was fixed to e=0. During the third visit of V1298 Tau, transits of both V1298 Tau d and b were observed. In this case, we modified the \verb|pycheops| transit fitting routine by modifying the default transit model to fit two transits within one light curve. For each planet, the orbital period, transit duration, impact parameter, and eccentricity are fixed. We fit the mid-transit time and transit depth of V1298 Tau b and d following a similar fitting approach as for the individual transits. 

The individual and joint raw CHEOPS observations of V1298 Tau c, d, and b along with the corresponding detrended light curves and best-fit models are shown in Figure \ref{fig:cheops_lcs} and Figure \ref{fig:cheops_lc_bd}. The planetary parameters used in the transit model, their priors and the best-fit values for each CHEOPS light curve are listed in Table \ref{tab:fit_parms}.

\section{Results}\label{sec:Res}

\subsection{High-precision radii and densities with \textit{CHEOPS}}

From high-precision CHEOPS transit observations of V1298 Tau, we report improved constraints on the transit depth measurements of planets c, d, and b, for which the corresponding best-fit values can be found in Table \ref{tab:fit_parms}. The CHEOPS measurements of $\rm R_{\mathrm{p}}/R_{\mathrm{s}}$ differ from those reported by \citet{sikora_2023} at the 2-3$\sigma$ level, while remaining consistent within 3$\sigma$ with the measurements reported by \citet{Feinstein_etal_2022}. The precision of the CHEOPS $\rm R_{\mathrm{p}}/R_{\mathrm{s}}$ measurements for V1298 Tau c, d, and b is improved by 30\%, 64\%, and 71\%, respectively, relative to \citet{sikora_2023}. The $\rm R_{\mathrm{p}}/R_{\mathrm{s}}$ measured from the joint transit of V1298 Tau d and b are within 1- and 4-$\sigma$ of that measured from the individual transits, respectively, with the 4-$\sigma$ difference due to a relatively larger error on the transit depth of V1298 Tau b within the joint transit.

For V1298 Tau c, we find a precision on $\rm R_{\mathrm{p}}/R_{\mathrm{s}}$ comparable to that achieved with TESS \citep{feinstein_2022}, despite the TESS measurements derived from 6 transits of the planet compared to a single CHEOPS transit. For V1298 Tau d and b, the precision on $\rm R_{\mathrm{p}}/R_{\mathrm{s}}$ is improved by 59\% and 74\%, respectively, compared to the TESS measurements. A summary of the $\rm R_{\mathrm{p}}/R_{\mathrm{s}}$ values and their associated uncertainties from this work and in literature is reported in Table \ref{tab:radiicompare}. We use the improved $\rm R_{\mathrm{p}}/R_{\mathrm{s}}$ measurements from CHEOPS to calculate the radii of V1298 Tau c, d, and b using the stellar radius from \citet{finociety_2023}, which assumes perpendicular stellar rotation axis to derive stellar parameters. 

Although JWST can achieve an order of magnitude better precision on $\rm R_{\mathrm{p}}/R_{\mathrm{s}}$ compared to CHEOPS, the uncertainty on the absolute planetary radius is fundamentally limited by the uncertainty on the stellar radius $\rm R_{\mathrm{s}}$. As a result, improvements in $\rm R_{\mathrm{p}}/R_{\mathrm{s}}$ beyond a certain level do not translate to proportionally tighter constraints on the radius of the planet. Within this context, the improved CHEOPS precision on $\rm R_{\mathrm{p}}/R_{\mathrm{s}}$ relative to previous optical measurements \citep[e.g.,][]{feinstein_2022} yields correspondingly tighter constraints on the planetary radius with the precision floor set by the stellar radius. We note that high-precision radius measurements from CHEOPS and JWST probe different atmospheric pressures and wavelength domains and are therefore complementary. 

\begin{figure}
    \centering
    \includegraphics[width=\columnwidth]{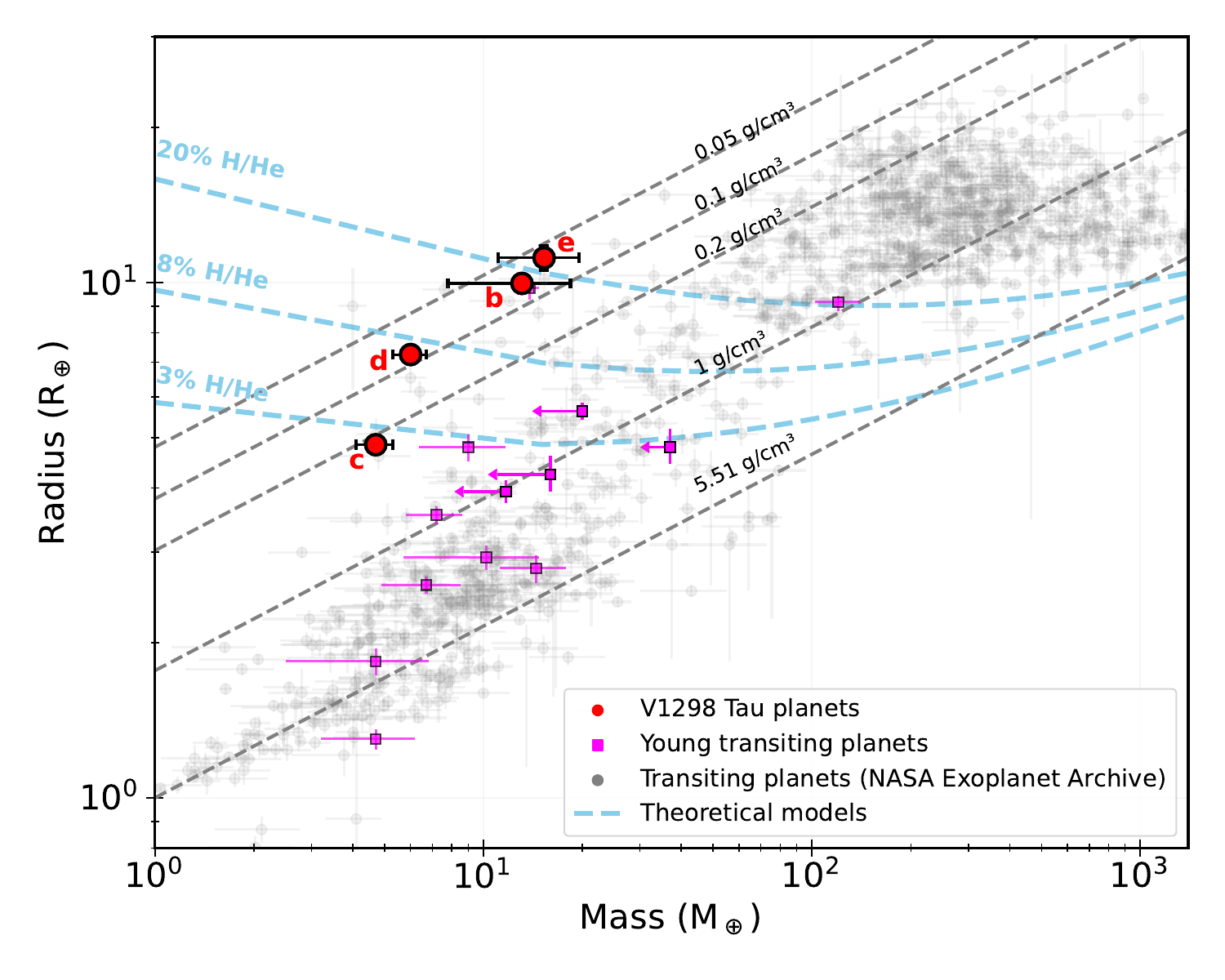}
    \includegraphics[width=\columnwidth]{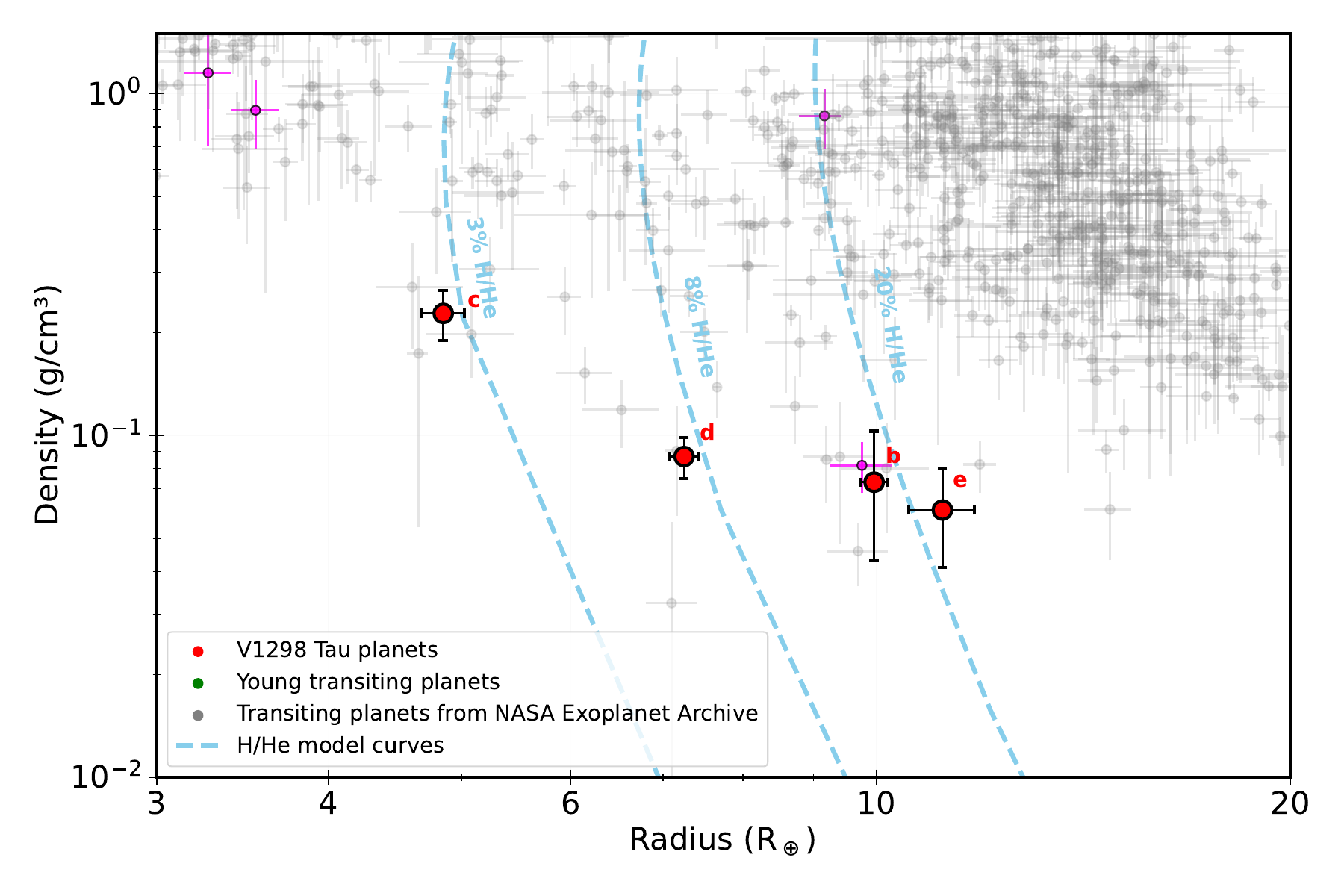}
    \caption{\textit{Top:} V1298 Tau in the mass-radius diagram of exoplanets. The V1298 Tau planets are indicated with red circles, where the radii of planets c, d, and b are from this work using CHEOPS. The radius of V1298 Tau e and masses of all planets are adopted from \citet{livingston_2026}. \textit{Bottom:} V1298 Tau in a density-radius diagram. Both figures show other known young transiting planets (magenta) for comparison. The population of mature transiting planets are shown in grey for comparison. The data are obtained from the NASA Exoplanet Archive. The skyblue dotted lines indicated theoretical models from \citet{lopezfortney_2014} calculated for different envelope fractions.}
    \label{fig:m-r_diagram}
    \vspace{0.5 cm}
\end{figure}

\begin{table*}
\centering
\caption{Radius, mass, and bulk density measurements of the V1298 Tau planets. The radii of the three innermost planets are derived in this work. The radius of V1298 Tau e and the masses of all planets are adopted from \citet{livingston_2026}.}
\label{tab:mass_rad_dens}
\begin{tabular}{lcccc}
\toprule
Planet & V1298 Tau c & V1298 Tau d & V1298 Tau b & V1298 Tau e \\
\midrule
Radius (R$_{\oplus}$) & $4.85 \pm 0.17$ & $7.25 \pm 0.18$ & $9.96 \pm 0.22$ & $11.17 \pm 0.61$ \\
Mass (M$_{\oplus}$) & $4.7 \pm 0.6$ & $6.0 \pm 0.7$ & $13.1 \pm 5.3$ & $15.3 \pm 4.2$ \\
Bulk density (g\,cm$^{-3}$) & $0.23 \pm 0.04$ & $0.09 \pm 0.01$ & $0.07 \pm 0.03$ & $0.06 \pm 0.02$ \\
\bottomrule
\end{tabular}
\end{table*}

\section{Discussion}\label{sec:disc}

\subsection{Bulk densities and implications on early evolution}
We use improved radii from this work and masses measured using TTV modeling \citep{livingston_2026} to calculate the bulk densities of V1298 Tau c, d, and b. For completeness, we also include the bulk density of V1298 Tau e from \citep{livingston_2026}. The densities of the V1298 Tau planets lie between 0.06-0.23 $\rm g/cm^3$. The innermost planet c is denser compared to the three outermost planet at a 3.4-$\sigma$ level. The densities of V1298 Tau d, b, and e are comparable with each other within the 1-$\sigma$ uncertainties reported. The radii, masses, and bulk densities of all four planets are summarized in Table~\ref{tab:mass_rad_dens}.

We present the mass-radius diagram of planets in the V1298 Tau system in the top panel of Fig. \ref{fig:m-r_diagram}. We include other known young planets ($<$300 Myr) in magenta for comparison. The mature exoplanet population is also shown in grey for comparison. The V1298 Tau planets are among the largest in radii in the known young planet population, particularly the two outermost planets b and e, lying between 10-11 R$\oplus$. We also present the density-radius diagram in the bottom panel of Fig. \ref{fig:m-r_diagram}. Among the known young planet population, the bulk density of planet b is comparable with that of HIP67522b. It should be noted that the mass of HIP67522b was measured using the scale height from atmospheric measurements \citep{Thao_2024}.

The low bulk densities of the V1298 Tau planets indicate that they could be retaining a substantial fraction of their envelopes at this stage, which is consistent with their young age and ongoing thermal contraction \citep{lopezfortney_2014}. As these planets evolve, their bulk densities are expected to increase due to subsequent cooling and contraction \citep{lopezfortney_2014, owen_2020, Kubyshkina_2022}, along with atmospheric mass loss driven by photoevaporation \citep{OwenWu_2013, Owen_2019}. In this context, the V1298 Tau planets are likely progenitors of the close-in sub-Neptune and potentially super-Earth populations observed around mature stars \citep{Fulton_2017, Poppenhaeger_2021}.

The relatively higher density of the innermost planet c suggests the planet retains a comparatively lower envelope mass fraction. This may indicate atmospheric escape shaping the atmospheric evolution of this planet and may place it in a regime where partial loss of its primordial H/He is likely. Similar signatures of ongoing atmospheric escape, including helium outflows, have been observed in other young systems, \citep[e.g. TOI-2076 b, c, d;][]{Zhang_2023, Wang_2026}. Although such processes cannot be directly inferred from bulk densities alone, dedicated atmospheric observations would be required to test whether H/He escape is currently ongoing in the V1298 Tau system.

\subsection{No evidence for present-day resonance and past excitation}\label{sec:early_evolution}

The V1298 Tau system is thought to have formed through disk-driven migration, which could have established a resonant chain involving the four planets. It has been proposed that the system may have formed one of two resonance chains: 3:2 -- 2:1 -- 2:1 or the 3:2 -- 2:1 -- 3:2 chain \citep{Tejada_etal_2022,Turrini_etal_2023}.

\citet{livingston_2026} show that the new period measurements place the system very close to a 3:2 -- 2:1 -- 2:1 resonance chain, with \(P_e/P_b = 2.0165\), \(P_b/P_d = 1.9466\), and \(P_d/P_c = 1.5034\), and infer a dynamically tranquil configuration based on the low eccentricities. Here, we further examine this interpretation of the dynamical state of the V1298 Tau system. We do this by conducting dynamical studies which assess: i) resonance angles, ii) forced eccentricities, and iii) the Normalized Angular Momentum Deficit (NAMD) of the system.

We performed an N-body integration using the \texttt{REBOUND} N-body integrator \citep{Rebound}, with the inbuilt symplectic integrator \texttt{WHFast}, and an integration timestep of $\num{1e-4}$ yr. We initialized the system containing the star and the four planets, with masses and eccentricities randomly sampled within uncertainties in Tab. \ref{tab:full_parms}. The orbital angles - mean anomaly ($M$), longitude of the pericentre ($\varpi$), and longitude of the ascending node ($\Omega$) - were randomly sampled between 0 and $2\pi$. The inclination of the planets was limited to be no larger than 1 degree, consistent with previous constraints \citep[e.g.][]{Turrini_etal_2023} and still allowing the possibility of resonances involving $\Omega$. We carry out 10\,000 integrations over 5\,000\,yr, sufficient to capture the secular precession of the planetary orbits. Within these integrations, we observed short term libration (some decades) of the resonance angles involving planets c and d in $\sim 6 \%$ of the initial parameters, likely due to their proximity to the exact commensurability. In all remaining cases, the resonance angles circulate rather than librate indicating no breach of resonance. We emphasize that while 10\,000 integrations constitute a substantial numerical sample, they cannot completely rule of resonant locking. The key result is that the system is dynamically stable regardless of whether resonance locking occurs.

We further investigate whether the observed eccentricities require past dynamical excitation by computing the forced eccentricities driven by mutual gravitational perturbations between the planets. We perform N-body integrations as described above, but fix the planetary masses to the mean value in Tab.~\ref{tab:mass_rad_dens} and initializing the system with zero eccentricity and inclination. The system is integrated over 5\,000 yr. We find that the forced eccentricities ($e^\mathrm{f}_\text{c} = \num{5.8e-3} \pm 0.006 $, $e^\mathrm{f}_\text{d} = \num{4.9e-3}{}^{+ 0.006}_{-0.005}$, $e^\mathrm{f}_\text{b} = \num{3.9e-3}  \pm 0.003$, and $e^\mathrm{f}_\text{e} = \num{9.7e-4} \pm 0.001$) are consistent with the observed values reported in \citet{livingston_2026}, demonstrating that the eccentricities arise purely from mutual planetary perturbations without invoking additional excitation mechanisms.

As an independent diagnostic, we compute the Normalized Angular Momentum Deficit (NAMD), a measure of the dynamical excitation on the system \citep{Chambers_2001}. Using masses and eccentricities sampled within their uncertainties in Tab.~\ref{tab:full_parms}, we obtain NAMD values ranging from $\num{2.0e-6}$ for planar orbits at the lower end of the mass range to $\num{2.2e-4}$ for an inclination of $1^\circ$ and upper mass limits. Even in the most extreme scenario, the NAMD is an order of magnitude lower than the Solar System's $\num{1.27e-3}$ value \citep{Turrini_et_al_2020}, indicating a low level of dynamical excitation.

\section{Conclusions}\label{sec:conc}

In this paper, we present new CHEOPS observations of the three innermost planets of the V1298 Tau system, yielding high precision measurements of the planetary radii ($\rm{R}_p/R_s$). CHEOPS achieves a level of precision comparable to that obtained from multiple TESS transits, despite CHEOPS requiring only a single transit per planet. These measurements significantly improve upon previous radius estimates and provide important constraints for comparative studies, including atmospheric characterization with \textit{JWST}.

Using updated radii from this work and TTV-derived masses from \citet{livingston_2026}, we derive revised bulk densities for the three innermost planets ranging between 0.06-0.23 $\rm g/cm^3$. We find that the innermost planet c is denser than the outermost planet at a 3.4-$\sigma$ level, while the bulk densities of the three outermost planets are consistent within the reported uncertainties. The relatively high density of V1298 Tau c could indicate more advanced atmospheric erosion where H/He escape may be ongoing, but bulk densities alone cannot confirm this. The V1298 Tau planets are some of the least dense planets among the known young planet population. Based on the calculated bulk densities, we can infer a gradient in envelope mass fraction from the innermost to the outermost planets, likely reflecting differential atmospheric evolution under strong stellar irradiation and ongoing thermal contraction within a single young system. In this context, the V1298 Tau planets represent early stage analogues of the sub-Neptune, and likely super-Earth, population observed around mature stars.

We further investigate the early dynamical state of the V1298 Tau system. From numerical integrations performed in this work, we find no evidence for present-day mean-motion resonance trapping. The forced eccentricities we calculate are in agreement with the observed eccentricities, implying they are primarily maintained by mutual gravitational perturbation rather than due to a dynamically excited past. Additionally, we find that the system maintains a low NAMD, an order of magnitude lower than that of the Solar System and comparable with planetary systems that have not experienced dynamically violent histories, such as the TRAPPIST-1 system, also consistent with a dynamically quiescent configuration. We note that the dynamically excited histories suggested by \citet{Turrini_etal_2023} are disfavored by the updated low masses, eccentricities, and orbital period measurements from \citet{livingston_2026}, which weaken the mutual perturbations and point to a more quiescent orbital architecture we recover here and consistent with the dynamical assessment of \citet{livingston_2026}.

\section*{Acknowledgments}

J.M.D acknowledges support from the Amsterdam Academic Alliance (AAA) Program, and the European Research Council (ERC) European Union’s Horizon 2020 research and innovation program (grant agreement no. 679633; Exo-Atmos). This work is part of the research program VIDI New Frontiers in Exoplanetary Climatology with project number 614.001.601, which is (partly) financed by the Dutch Research Council (NWO). PM acknowledges support from STFC research grants ST/R000638/1, ST/Y002563/1 and UKRI1193 and UK Space agency grant UKRI966. V.P. would like to thank support from the UKRI STFC grant UKRI1171.

\section*{Data Availability}
All observational data analyzed in this work are publicly available on the CHEOPS archive.
 



\bibliographystyle{mnras}
\bibliography{sample} 




\appendix


\section{Data reduction: DRP vs. PIPE}

\begin{figure*}
    \centering
    \includegraphics[width=0.8\textwidth]{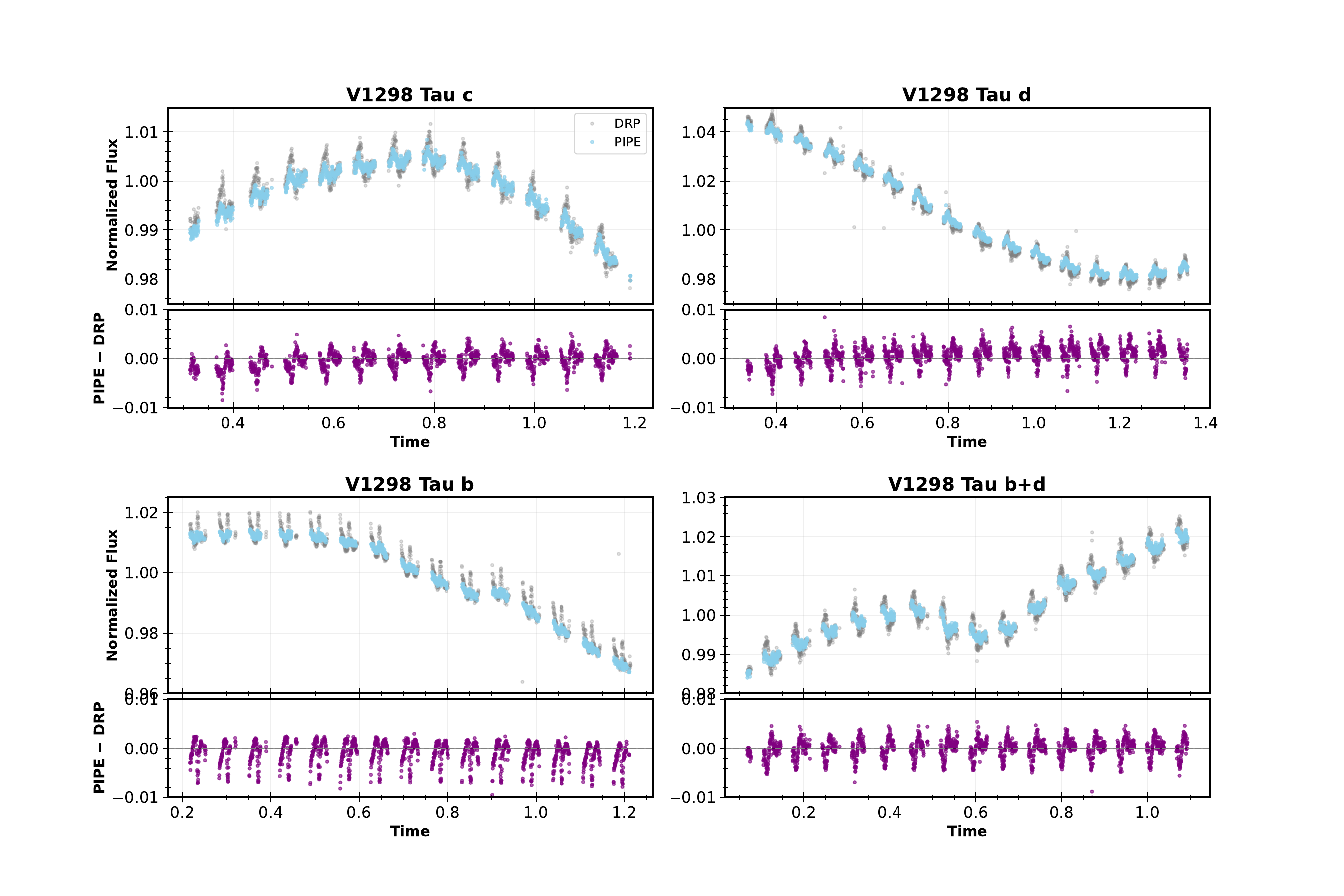}
    \caption{\textit{CHEOPS} transit observations of V1298 Tau c, d, b, and b+d extracted using PSF photometry (PIPE; in blue) and aperture photometry (DRP; in grey) pipelines. For each planet, the top row presents the raw (pre-detrended) light curves extracted using PIPE (skyblue) and DRP (grey). The bottom panel shows the difference between the PIPE and DRP extracted light curves.}
    \label{fig:pipevsdrp}
\end{figure*}

Since \textit{CHEOPS} operates in a nadir-locked orbit around the Earth, its field of view rotates about the target with a period of approximately 99 minutes. This rotation introduces flux modulation correlated with the telescope roll angle. The flux modulation is amplified when variable contamination is introduced by nearby stars entering and exiting the photometric aperture. Point-spread function (PSF) mitigates this effect by modeling and subtracting the contributions of background stars, thereby reducing roll-angle dependent flux variations.

The CHEOPS field of view surrounding V1298 Tau is relatively crowded, containing several bright background sources. The PSF of V1298 Tau is primarily contaminated by the extended PSF wings of a fainter star (\textit{Gaia} DR3 51886331671984640; m$_{\rm Gaia} \simeq$ 16.8824). To account for this crowded field, we extracted light curves using the open source photometric package PIPE \citep{Brandeker_2024}. 
Following recommendations for targets with $G_{\rm mag} < 11$, we adopted a "klip" size of one for the principal component bases. The static image was not removed since the observations consisted of a single subarray and no imagettes. All other processing steps, including fitting and subtracting the PSF of background stars, removing the flat and dark field, masking bad pixels, and removing any satellite streaks were performed using the default PIPE settings.

We compared the CHEOPS light curves of the V1298 Tau planets extracted using PIPE and the standard Data Reduction Pipeline (DRP), as shown in Figure \ref{fig:pipevsdrp}. For the V1298 Tau c, d and b+d observations, PIPE photometry yields lower median absolute deviation (MAD; by 3-5 \%), reduced flux uncertainties (by $\sim$10 \%) and lower RMS (by $\sim$10 \%) compared to the DRP extracted light curves. For V1298 Tau b, the flux uncertainty and RMS are consistent between PIPE and DRP, while the MAD is lower for PIPE (11508 ppm) compared to DRP (11066 ppm). We adopt the PIPE extracted light curves for our analysis as our primary objective for light curve extraction is to minimize the contamination driven flux modulation caused by the nearby background stars over the course of the observations.

\section{Light curve fitting}

\begin{figure}
    \centering
    \includegraphics[width=0.6\columnwidth]{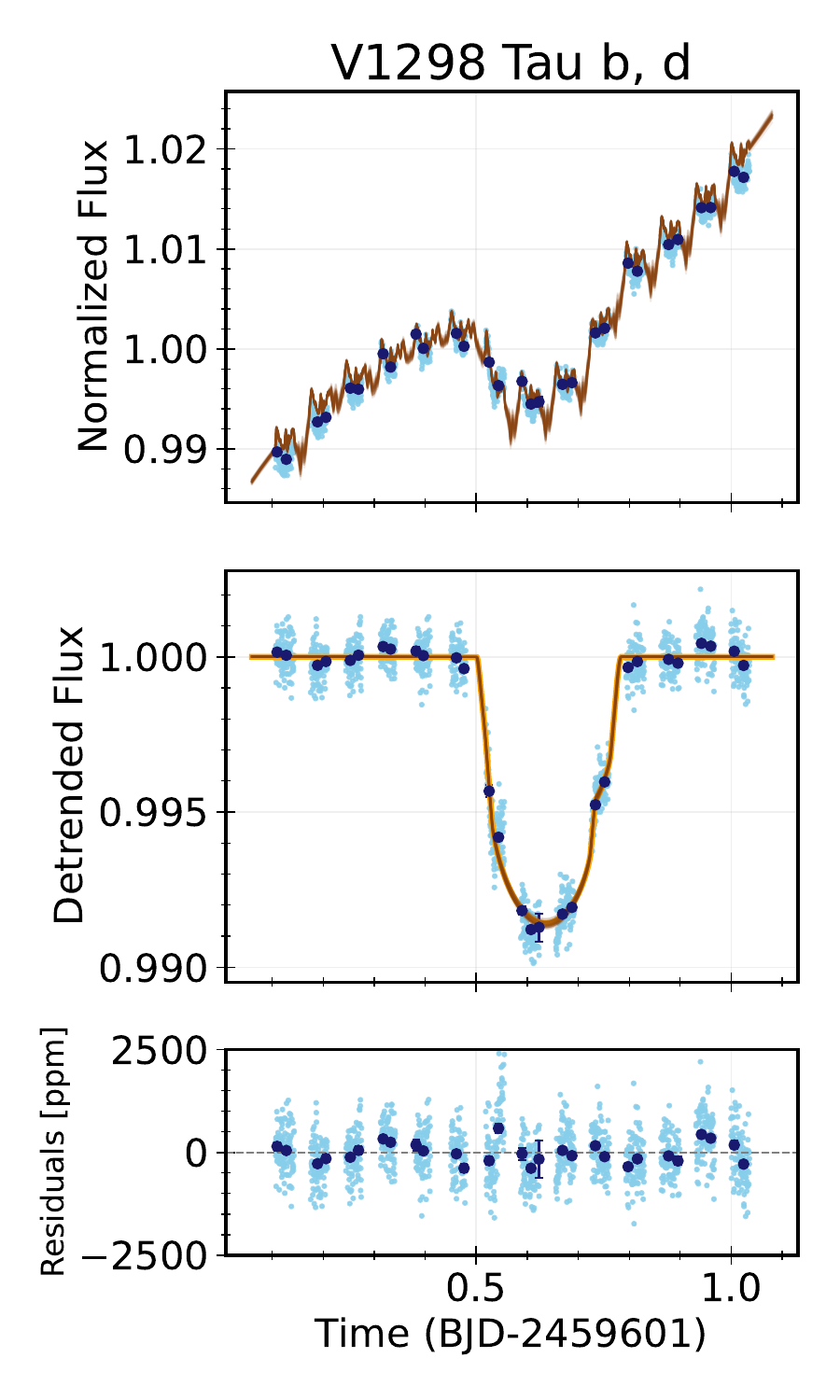}
    \caption{Overlapping transits of V1298 Tau b and d measured by CHEOPS. The top panel shows the raw transit light curve with best fit model overplotted in maroon. The middle panel shows the normalized transit light curve with both planets, with best fit normalized models overplotted as continuous brown curves. The residuals are shown in the bottom panel.}
    \label{fig:cheops_lc_bd}
    \vspace{0.5 cm}
\end{figure}

\begin{table*}
\centering
\caption{Decorrelation parameters used in the systematics model to fit CHEOPS observations of V1298 Tau.}
\label{tab:decorrparms}
\begin{tabular}{lcl}
\toprule
Visit & Transit(s) Observed & Detrending parameters \\
\midrule
1 & V1298 Tau d & bg, y, t, t$^{2}$, t$^{3}$, t$^{4}$, cos$\phi$, sin$2\phi$, cos$2\phi$, cos$3\phi$ \\
2 & V1298 Tau c & bg, y$^{2}$, t, t$^{2}$, t$^{3}$, sin$\phi$, cos$\phi$, sin$2\phi$, sin$3\phi$, cos$3\phi$ \\
3 & V1298 Tau b, d & bg, t, t$^{2}$, t$^{3}$, t$^{4}$, t$^{5}$, sin$\phi$, cos$\phi$, sin$2\phi$, sin$3\phi$, cos$3\phi$ \\
4 & V1298 Tau b & bg, t, t$^{2}$, t$^{3}$, t$^{4}$, t$^{5}$, sin$\phi$, cos$\phi$, sin$2\phi$, cos$2\phi$, sin$3\phi$, cos$3\phi$ \\
\bottomrule
\end{tabular}
\end{table*}

\begin{table*}
\centering
\caption{Transit depth and its respective precision for each of the first three planets of V1298 Tau measured in this work and compared with previous works (i.e. \citet{david_2019b, feinstein_2022, Suarez_etal_2022, sikora_2023, livingston_2026}).}
\label{tab:radiicompare}
\begin{tabular}{lcccccc}
\toprule
 & This work & \citet{david_2019b} & \citet{feinstein_2022} & \citet{Suarez_etal_2022} & \citet{sikora_2023} & \citet{livingston_2026}  \\
\midrule
\multicolumn{7}{c}{V1298 Tau c} \\
\midrule
R$_p$/R$_s$ & 0.03106 & 0.0381 & 0.0337 & 0.0371 & 0.0354 & 0.0356 \\
Precision   & $\pm$0.00090 & $\pm$0.0017 & $\pm$0.0009 & $\pm$0.0019 & $\pm$0.0013 & $^{+0.0018}_{-0.0015}$\\
\midrule
\multicolumn{7}{c}{V1298 Tau d} \\
\midrule
R$_p$/R$_s$ & 0.04647 & 0.0436 & 0.0409 & 0.0464 & 0.0429 & 0.0458\\
Precision   & $\pm$0.00062 & $^{+0.0024}_{-0.0021}$ & $^{+0.0014}_{-0.0015}$ & $\pm$0.0020 & $^{+0.0016}_{-0.0017}$ & $\pm$0.0015 \\
\midrule
\multicolumn{7}{c}{V1298 Tau b} \\
\midrule
R$_p$/R$_s$ & 0.06386 & 0.0700 & 0.0636 & 0.0698 & 0.0673 & 0.0661\\
Precision   & $\pm$0.00049 & $\pm$0.0023 & $\pm$0.0018 & $\pm$0.0024 & $\pm$0.0017 & $^{+0.0019}_{-0.0017}$\\
\midrule
\multicolumn{7}{c}{V1298 Tau e} \\
\midrule
R$_p$/R$_s$ & -- & 0.0611 & 0.0664 & 0.0583 & 0.0643 & 0.0716 \\
Precision   & -- & $^{+0.0052}_{-0.0037}$ & $^{+0.0025}_{-0.0021}$ & $\pm$0.0040 & $\pm$0.0029 & $\pm$0.0036\\
\bottomrule
\end{tabular}
\end{table*}

\begin{figure*}
    \centering
    \includegraphics[width=\textwidth]{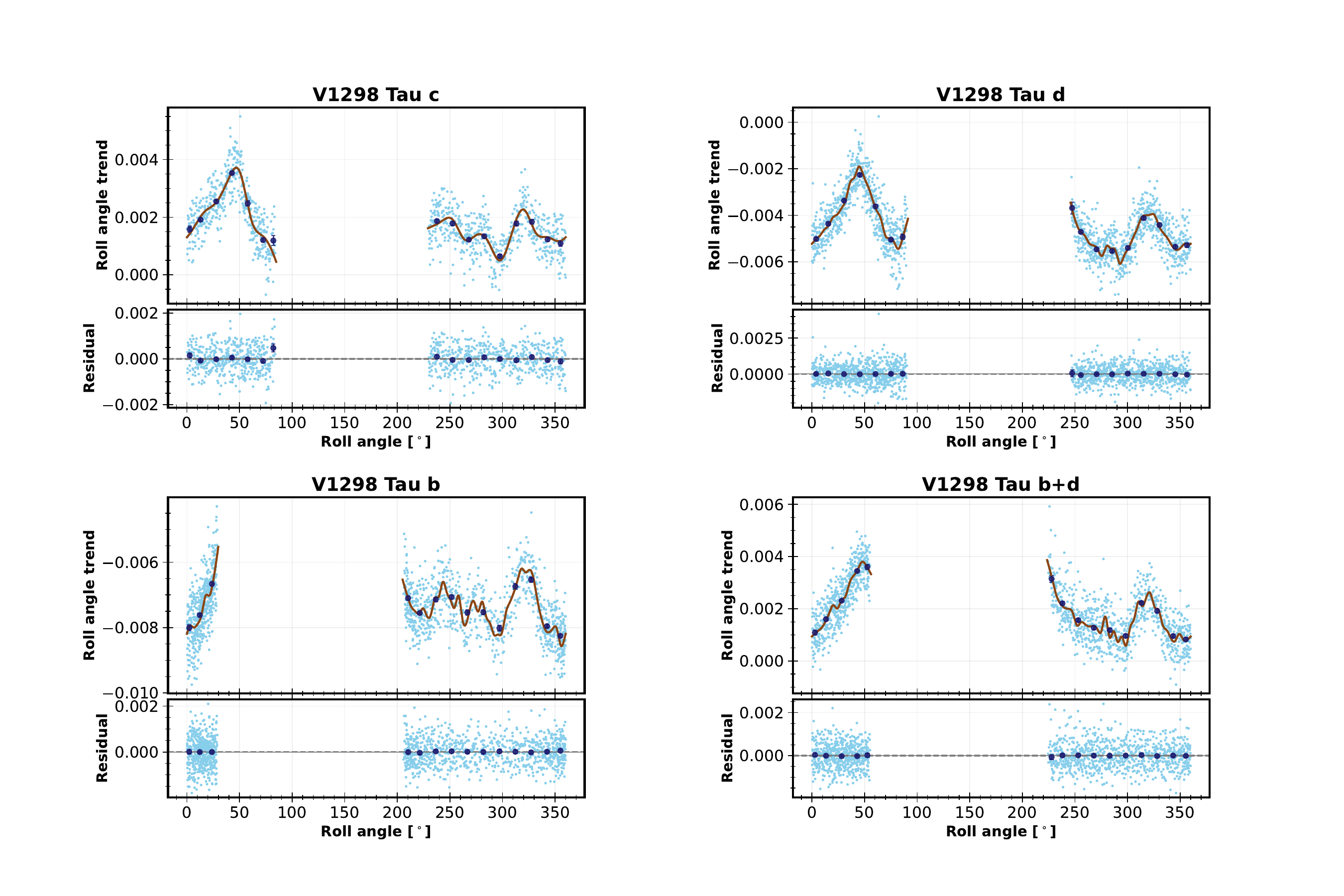}
    \caption{Trends in CHEOPS data of V1298 Tau c, d, b, and b+d as a function of roll angle. For each observation, the top panel shows the residuals from the best-fit transit model (blue) and the best-fit roll angle model (brown) as a function of roll-angle. The bottom panel shows the residuals from the best-fit model including roll-angle trends.}
    \label{fig:rollangle}
    \vspace{0.5 cm}
\end{figure*}

\begin{table*}
\centering
\caption{Planetary parameters of the V1298 Tau planets used for the dynamical analysis of the current paper. Planet-to-star radius ratios (Rp/Rs) measured in this work are converted to planetary radii using stellar radius value of Rs = 1.43 ± 0.03 R$_{\odot}$ from \citet{finociety_2023}. The Rp/Rs of V1298 Tau e reported in Extended Table 1 of \citet{livingston_2026} is converted to planetary radius using stellar radius value from \citet{finociety_2023}. Planetary mass, eccentricity, and period of all planets are adopted from Table 1 of \citep{livingston_2026}. The semi-major axis of V1298 Tau e reported in \citet{livingston_2026} is converted from units of AU to a/Rs using stellar radius from \citet{finociety_2023}. For the dynamical analysis, we assume a stellar mass value of Ms = 1.26±0.06 M$_{\odot}$ and stellar rotation period 2.910±0.005 days \citep{finociety_2023}. The eccentricities are reported in percent.} 
\label{tab:full_parms}
\begin{tabular}{lcccccc}
\toprule
Parameter & Unit & V1298 Tau c & V1298 Tau d & V1298 Tau b & V1298 Tau e & Comments \\
\midrule
Radius & R$_{\oplus}$ & $4.85 \pm 0.17$ & $7.25 \pm 0.18$ & $9.96 \pm 0.22$ & $11.17 \pm 0.61$ & V1298 Tau c, d, b: this work \\
Mass & M$_{\oplus}$ & $4.7 \pm 0.6$ & $6.0 \pm 0.7$ & $13.1 \pm 5.3$ & $15.3 \pm 4.2$ & - \\
Eccentricity & \% & <0.94 & <0.87 & 0.79 ± 0.41 & <1.24 & - \\
Period & days & 8.249164(3) & 12.401394(9) & 24.140006(17) & 48.677714(53) & - \\
Semi-Major axis & a/R$_s$ & $13.34\pm0.012$ & $17.42\pm0.011$ & $27.16\pm0.02$ & $40.44\pm 1.04$ & V1298 Tau c, d, b: this work \\
\bottomrule
\end{tabular}
\end{table*}


\bsp	
\label{lastpage}
\end{document}